\documentclass[epj,referee]{svjour}
\usepackage{amssymb}
\usepackage{latexsym}
\usepackage{graphics}
\begin{document}
\title{
Duality-mediated
critical amplitude ratios for
 the $(2+1)$-dimensional $S=1$
$XY$ model
}
\author{Yoshihiro Nishiyama  
}                     
\offprints{}          
\institute{Department of Physics, Faculty of Science,
Okayama University, Okayama 700-8530, Japan}
\date{Received: date / Revised version: date}
%
\abstract{
The phase transition for
the $(2+1)$-dimensional
spin-$S=1$ $XY$ model was investigated numerically.
Because of the boson-vortex duality,
the spin stiffness $\rho_s$
in the ordered phase
and the vortex-condensate stiffness $\rho_v$
in the disordered phase
should have a close relationship.
We employed the exact diagonalization method,
which yields the excitation gap
directly.
As a result,
we estimate the
amplitude ratios $\rho_{s,v}/\Delta$
($\Delta$: Mott insulator gap)
by means of the scaling analyses
for the finite-size cluster with $N \le 22$ spins.
The ratio $\rho_s/\rho_v$
admits 
a quantitative measure 
of 
deviation from selfduality.
} 
\maketitle

\section{\label{section1}Introduction}

The $(2+1)$-dimensional boson system undergoes
the superfluid-insulator transition,
which belongs to the three-dimensional 
$XY$ universality class.
The system possesses
the boson-vortex duality \cite{Stone78,Fisher89,Wen90},
which brings about rich characters as to the critical phenomenon.
The superfluid (ordered) and insulator (disordered)
phases are characterized by 
the spin-  and 
vortex-condensate-stiffness 
constants $\rho_{s,v}$, respectively.
The critical amplitude ratio
$\rho_s/\rho_v$ admits a ``quantitative measure'' \cite{Gazit14}
of deviation from selfduality.
Note that each stiffness constant
$\rho_{s,v}$ yields the Drude weight
for the corresponding AC conductivity
\cite{Swanson14},
and
it may be accessible \cite{Gazit14}
experimentally \cite{Corson99,Crane07,Sherson10}.
In this sense, the duality-mediated amplitude ratio
is not a mere theoretical concept.

In Fig. \ref{figure1},
we present a schematic drawing 
for the
stiffness constants
(order parameters) $\rho_{s,v}$,
which develop
in the 
superfluid ($\delta J_{NN}>0$) and 
insulator ($\delta J_{NN}<0$) phases, respectively.
In the respective phases,
there open
the Higgs-mass $m_H$ and
the
Mott-insulator $\Delta$ gaps above the ground state;
in the terminology of the quantum  spin model, the former (latter)
corresponds to the longitudinal mode of the magnetic moment
(paramagnetic massive excitation).
(On the one hand, the Goldstone excitation
corresponds to the transverse mode of the magnetic moment.)
A key ingredient is that
these quantities possess the
identical scaling dimension,
and the amplitude ratios among them make sense.

Recently,
the
duality-mediated
amplitude
ratios $\rho_{s,v}/\Delta$ came under theoretical 
investigation
by means of
the  Monte Carlo 
\cite{Gazit14,Gazit13}
and
renormalization-group 
\cite{Rose17,Rose15,Ranson13} methods.
In prior to these studies,
the mass-gap amplitude ratio
$m_H / \Delta $ has been investigated rather extensively 
 \cite{Gazit13,Gazit13a,Chen13,Nishiyama15,%
Rancon14,Katan15}. 
As a matter of fact,
for the two-dimensional ultra cold atom,
these mass gaps
were observed in proximity to the superfluid-insulator
transition
\cite{Endres12};
see Ref. \cite{Pekker15} for a review.

The aim of this paper is to investigate these amplitude ratios
for
the $(2+1)$-dimensional $S=1$ $XY$ model (\ref{Hamiltonian})
by means of the exact diagonalization method.
This method allows us to calculate the energy gaps
such as
$m_H$ and $\Delta$
without resorting to the inverse Laplace transformation \cite{Gazit13}.
Moreover, the 
ground-state resolvent (response function)
such as the second term of
Eq. (\ref{spin_stiffness}) 
is evaluated by means of the continued-fraction-expansion method
\cite{Gagliano87};
the continued-fraction-expansion method
is essentially the same as the Lanczos 
tri-diagonalization algorithm,
and computationally less demanding.
Otherwise, the finite-temperature effect
(anisotropy between the imaginary-time and real-space
system sizes)
has to be considered carefully
\cite{Gazit14s}.

The Hamiltonian
for the two-dimensional $S=1$ $XY$ model
is given by
\begin{eqnarray}
{\cal H} &=&
 -J_{NN}\sum_{\langle ij\rangle}(S^x_iS^x_j+S^y_iS^y_j)-
J_{NNN}\sum_{\langle\langle ij\rangle\rangle}(S^x_iS^x_j+S^y_iS^y_j) \nonumber \\
\label{Hamiltonian}
 & & + D_{\Box}\sum_{[ijkl]} (S^z_i+S_j^z+S^z_k+S^z_l)^2
  +D\sum_{i=1}^N(S^z_i)^2 ,
\end{eqnarray}
with the $S=1$-spin operators $\{ {\bf S}_i \}$
placed at the square-lattice points, $i=1,2,\dots , N$.
Here,
the summations 
$\sum_{\langle ij \rangle,\langle\langle ij \rangle\rangle,[ijkl]}$
run over all possible nearest-neighbor 
$\langle ij \rangle$, 
next-nearest-neighbor
$\langle \langle ij \rangle\rangle$, and 
plaquette spins
$[ijkl]$, respectively.
The parameters $J_{NN,NNN}$ and $D_\Box$ are the corresponding
coupling constants.
The symbol
$D$ denotes the single-ion anisotropy.
We restrict ourselves to the parameter subspace
\begin{eqnarray}
	(J_{NN},J_{NNN},D_\Box,D) &=&
	(J_{NN}^*,J_{NNN}^*,D_\Box^*,D^*)  \nonumber \\
 & & +
(\delta J_{NN},\delta J_{NNN},\delta D_\Box,0)
,
\end{eqnarray}
beside the critical point \cite{Nishiyama08}
\begin{eqnarray}
	 & &
(J_{NN}^*,J_{NNN}^*,D_\Box^*,D^*)
  =
    \nonumber \\
\label{FP_intro}
& & 
(
 0.158242810 160,
0.058561393564,
0.10035104389  ,
0.957)  .
\end{eqnarray}
The critical point (\ref{FP_intro})
was fixed \cite{Nishiyama08}
so as to suppress 
corrections to scaling;
namely, the critical point was determined via
a preliminary approximate real-space-decimation method
followed by
the (ordinary) finite-size-scaling scheme.

The rest of this paper is organized as follows.
In the next section, we present the simulation results.
A comparison with the preceeding results is made as well.
In Sec. \ref{section3},
we address the summary and discussions.

\section{\label{section2}
Numerical results}

In this section, we present the numerical results.
We employed the exact diagonalization method
for the two-dimensional $S=1$ $XY$ model (\ref{Hamiltonian}).
We implemented the screw-boundary condition \cite{Novotny90},
which enables us to treat a variety of system sizes $N=10,12,\dots,22$;
the simulation algorithm is presented in Ref. \cite{Nishiyama08}.
Because the $N$ spins constitute a rectangular cluster,
the linear dimension of the cluster is given by $L=\sqrt{N}$,
which sets a fundamental  length
scale for the scaling analyses.

\subsection{\label{section2_1}
Scaling behavior of the spin stiffness $\rho_s$:
Analysis of $\rho_s/m_H$}

In this section, we investigate the scaling behavior 
for the spin stiffness 
(superfluid density) $\rho_s$
(\ref{spin_stiffness}).
We 
look into the
interaction subspace
\begin{equation}
(\delta J_{NN},\delta J_{NNN} , \delta D_{\Box})
=
\left(
	\delta J_{NN}, \frac{J^*_{NNN}\delta J_{NN}}{J^*_{NN}},0
	 \right)
,	 
\end{equation}
 parameterized by 
$\delta J_{NN}$.
In a preliminary survey,
we found that 
within this interaction subspace,
the ratio $\rho_s/m_H$ is kept 
invariant
for a considerably wide range of $\delta J_{NN}$.

In Fig. \ref{figure2},
we present the scaling plot,
$\delta J_{NN} L^{1/\nu}$-$L \rho_s(\delta J_{NN})$,
for various 
($+$) $N=18$,
($\times$) $20$, and
($*$) $22$.
Here,
the scaling parameter $\nu$
is set to
$\nu=0.6717$,
which is
taken from the existing literatures, Ref.
\cite{Campostrini06,Burovski06};
note that the superfluid-insulator
criticality belongs to the
three-dimensional $XY$ universality class.
Hence, there is no adjustable fitting parameter involved in the
present scaling analysis.
The spin stiffness is given by the formula
\begin{equation}
	\label{spin_stiffness}
\rho_s = \frac{1}{N} \langle 0 | T | 0 \rangle
        +\frac{2}{N} \left\langle 0 
	\left| J 
	\frac{{\cal P}}{
	{\cal H}-E_0}
	J
	\right| 0 \right\rangle
       ,
\end{equation}
with the ground-state energy (vector) 
$E_0$ ($|0\rangle$),
projection operator ${\cal P}=1-|0\rangle\langle0|$,
the current operator $J$,
and
the
 diamagnetic contribution $T$;
explicit expressions for $J$ and $T$
are presented in Appendix.
We stress that the 
{\em ground state} resolvent 
(the second term of Eq. (\ref{spin_stiffness}))
is readily evaluated 
with the continued-fraction expansion \cite{Gagliano87}.

\begin{figure}
\resizebox{0.5\textwidth}{!}{%
\includegraphics{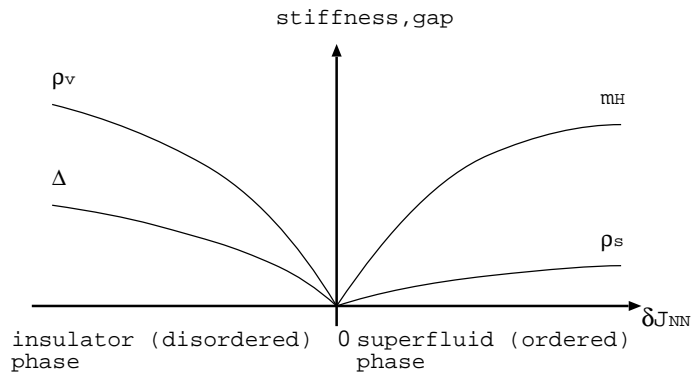}}%
\caption{
\label{figure1}
The spin-
and vortex-condensate-stiffness
constants $\rho_{s,v}$
characterize the 
superfluid ($\delta J_{NN} >0$)
and insulator ($\delta J_{NN}<0$) phases, respectively.
Correspondingly,
the Higgs-mass $m_H$ and 
the Mott-insulator $\Delta$ gaps 
open above the ground state.
These quantities have the same scaling dimension,
and the critical amplitude ratios among them
make sense.
The ratio $m_H/\Delta$ has been
investigated rather extensively so far
\cite{Gazit13a,Chen13,Nishiyama15,Rancon14,Katan15}.
}
\end{figure}

\begin{figure}
\includegraphics{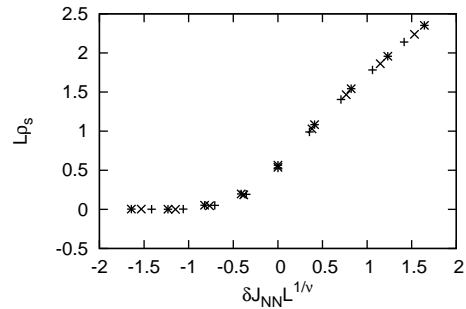}%
\caption{
\label{figure2}
The scaling plot,
$ \delta J_{NN} L^{1 / \nu}$-$L \rho_s$,
is presented for various system sizes 
($+$) $N=18$, 
($\times$) $20$, and
($*$) $22$. 
Here, the scaling parameter 
$\nu=0.6717$ is taken from the 
existing literatures 
\cite{Campostrini06,Burovski06};
namely,
there is no adjustable parameter involved in the scaling analysis.
The spin stiffness $\rho_s$
develops in the superfluid phase,
$\delta J_{NN} >0$.
}
\end{figure}

The data in Fig. \ref{figure2}
appear to collapse into a scaling function satisfactorily,
indicating that the simulation result
reaches the scaling regime.
As mentioned in Introduction,
corrections to scaling are suppressed by
finely adjusting 
\cite{Nishiyama08}
the interaction parameters
to Eq. (\ref{FP_intro}).
As shown in Fig. \ref{figure1}.
The spin stiffness $\rho_s$
increases (decreases)
in the superfluid (insulator) phase;
see Fig. \ref{figure1} as well.
The scaling plot in Fig. \ref{figure2}
indicates that the ordinate $L\rho_s$
is dimensionless.
That is,
the stiffness constant possesses the scaling dimension
of either reciprocal correlation length or excitation gap.
Therefore, it is expected that the ratio $\rho_s / m_H$ 
should be a universal constant.

\begin{figure}
\includegraphics{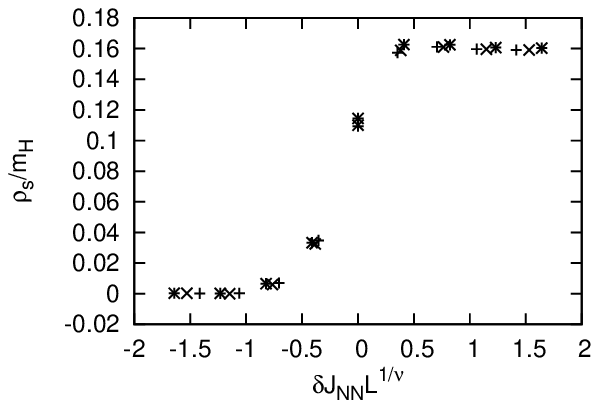}%
\caption{
\label{figure3}
The scaling plot,
$ \delta J_{NN} L^{1/\nu}$-$\rho_s(\delta J_{NN})/m_H(\delta J_{NN})$,
is presented for various system sizes
($+$) $N=18$,
($\times$) $20$,
($*$) $22$;
the scaling parameter $\nu$ is the same as that of Fig. 
\ref{figure2}.
A plateau extends around $\delta J_{NN} L^{1/\nu} >0.5 $,
suggesting that the amplitude ratio 
$\rho_s/m_H$
takes a constant value.
}
\end{figure}

Stimulated by this observation,
we turn to the analysis of 
the amplitude ratio $\rho_s/m_H$.
In Fig.  \ref{figure3},
we present the scaling plot,
$\delta J_{NN} L^{1/\nu}$-$\rho_s(\delta J_{NN})/m_H(\delta J_{NN})$,
for 
($+$) $N=18$,
($\times$) $20$, and
($*$) $22$;
the scaling parameter $\nu$ is the same as that of Fig. 
\ref{figure2}.
The Higgs mass $m_H$
is evaluated by the
formula
$m_H = E_1 - E_0$
with
the first-excitation energy $E_1$ 
specified by the 
 the zero-momentum 
($k=0$)
and
zero-total-magnetization ($S^z=0$) indices;
the Goldstone mode belongs to the Hilbert-space sector
with $S^z= \pm 1$ and $k=0$.
(In this way, we are able to estimate the excitation gaps
in a straightforward manner without resorting to
the inverse Laplace transformation \cite{Gazit13}.)
A plateau extends around $\delta J_{NN}L^{1/\nu} > 0.5 $.
Such a feature 
indicates that
 the amplitude ratio 
$\rho_s / m_H$ takes a constant value
in proximity to the critical point.

\begin{figure}
\includegraphics{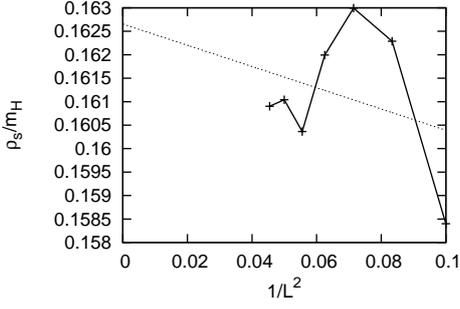}%
\caption{
\label{figure4}
The approximate
amplitude ratio 
$\rho_s/m_H$  
(\ref{appr1})
is plotted for 
$1/L^2$.
The least-squares fit to these data
yields an estimate 
$\rho_s/m_H=0.1627(23)$ in the thermodynamic limit.
The irregularities (bumps) around
$1/L^2=0.05(\approx 1/4.5^2)$ and $0.08(\approx 1/3.5^2)$ are due to the 
artifact of the screw-boundary 
condition
\cite{Novotny90}.
A possible systematic error is considered in the text.
}
\end{figure}

In Fig. \ref{figure4},
we plot the approximate amplitude ratio $\rho_s/m_H$ for $1/L^2$.
Here, the approximate amplitude ratio 
$\rho_s / m_H $
denotes the plateau height
\begin{equation}
\label{appr1}
\left.
\frac{\rho_s (\delta J_{NN})}{m_H(\delta J_{NN})}
\right|_{\delta J_{NN}=\delta \bar{J}_{NN}},
\end{equation}
at the extremal point,
$\partial_{\delta J_{NN}} (\rho_s/m_H) |_{\delta J_{NN}=\delta \bar{J}_{NN}} =0$, 
for each system size $N(=L^2)=10,12,\dots,22$.
The least-squares fit to these data yields an estimate
$\rho_s/m_H=0.1627(23)$ 
in the thermodynamic limit $L\to\infty$.
In order to appreciate a possible systematic error,
we make an alternative extrapolation.
The irregularities (bumps) around
$1/L^2=0.05(\approx 1/4.5^2)$ and $0.08(\approx 1/3.5^2)$ are due to the 
artifact of the screw-boundary 
condition
\cite{Novotny90}.
In order to avoid these intermittent irregularities,
we consider a sector $N=14,16,\dots ,20$,
for
which 
the least-squares fit
leads to
 an estimate
$\rho_s/m_H=0.1551(25)$.
The discrepancy $\approx 0.008$ 
between the independent extrapolations may indicate a possible systematic error.
Putting both
least-squares-fit and systematic errors
into consideration,
we estimate the amplitude ratio
as
\begin{equation}
\label{estimate1}
\rho_s / m_H =0.16(1)
.
\end{equation}

A remark is in order.
According to our preliminary survey,
the combination 
$\rho_s / m_H$ turned out to be an optimal
one
in the sense that the ratio exhibits
a stable plateau for a considerably wide range of $\delta J_{NN}$.
Experimentally,
the Higgs mass
 $m_H$ became observable
 even in close vicinity of the critical point
\cite{Endres12}.

\subsection{\label{section2_2}
Scaling behavior of 
the vortex-condensate stiffness $\rho_v$:
Analysis of $\rho_s/\rho_v$}

In this section,
we analyze the amplitude ratio 
$\rho_s (\delta J_{NN}) / \rho_v(-\delta J_{NN})$.
For that purpose,
we survey the parameter subspace
\begin{eqnarray}
& & (\delta J_{NN},\delta J_{NNN}, \delta D_\Box) \nonumber \\
&=&
\left(
	 \delta J_{NN},\frac{J_{NNN}^* \delta J_{NN}}{J_{NN}^*},\frac{1.4
D^*_{\Box} \delta J_{NN}}{J^*_{NN}}
 \right)
     ,
\end{eqnarray} 
parameterized by a single
variable 
$\delta J_{NN}$.
In a preliminary survey,
we found that 
within this subspace,
 the amplitude ratio
$\rho_s/\rho_v$ exhibits a stable plateau in 
an appreciable range of $\delta J_{NN}$.

\begin{figure}
\includegraphics{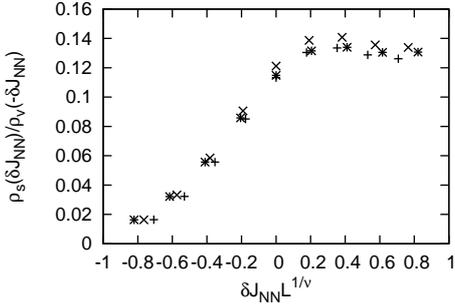}%
\caption{
\label{figure5}
The scaling plot,
$ \delta J_{NN} L^{1/\nu}$-$\rho_s(\delta J_{NN})/\rho_v(-\delta J_{NN})$,
is presented for various system sizes
($+$) $N=18$,
($\times$) $20$,
and 
($*$) $22$;
the scaling parameter $\nu$ is the same as that of Fig. 
\ref{figure2}.
A plateau extends around 
$\delta J_{NN} L^{1/\nu} > 0.2 $,
suggesting that the amplitude ratio 
$\rho_s/\rho_v$
takes
a universal constant.
}
\end{figure}

In Fig. \ref{figure5}, we present the scaling plot,
$\delta J_{NN} L^{1/\nu}$-$\rho_s(\delta J_{NN})/\rho_v(-\delta J_{NN})$, 
for 
($+$) $N=18$,
($\times$) $20$,
and 
($*$) $22$;
the scaling parameter $\nu$ 
is the same as that of Fig. \ref{figure2}.
Here, the vortex-condensate stiffness $\rho_v$
is given
\cite{Gazit14}  
by
$\rho_v = (2 \pi)^{-2} k_1^2/ \chi_\rho(k_1) $
with 
the 
charge-density-wave susceptibility $\chi_\rho(k)$
and
$k_1=2\pi/N$;
the explicit expression for $\chi_\rho$
is presented in Appendix.
The plateau 
extending around $\delta J_{NN} L^{1/\nu} > 0.2  $ indicates
that
the amplitude ratio $\rho_s/\rho_v$
is indeed a universal constant.

\begin{figure}
\includegraphics{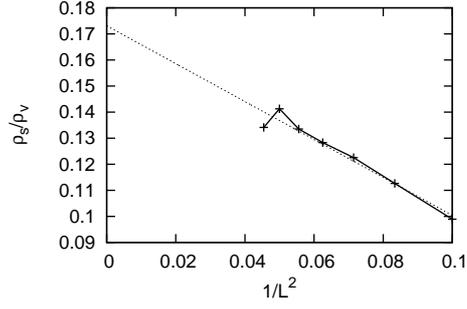}%
\caption{
\label{figure6}
The approximate
amplitude ratio 
$\rho_s / \rho_v$
(\ref{appr2}) 
is plotted for $1/L^2$.
The least-squares fit to these data
yields an estimate 
$\rho_s/\rho_v=0.1731(50)$ in the thermodynamic limit.
The irregularity around 
$1/L^2=0.051(\approx 1/4.5^2)$
is an artifact due to the screw-boundary condition \cite{Novotny90}.
A possible systematic error is considered in the text.
}
\end{figure}

In Fig. \ref{figure6},
we present the approximate ratio 
$\rho_s/\rho_v$
for $1/L^2$.
Here, the approximate ratio 
$\rho_s /\rho_v$
denotes the plateau height
\begin{equation}
\label{appr2}
 \left.
\frac{\rho_s (\delta J_{NN})}
     {\rho_v (-\delta J_{NN})}
     \right|_{\delta J_{NN}=\delta \tilde{J}_{NN}}
,
\end{equation}
with
$\partial_{\delta J_{NN}} ( \rho_s/ \rho_v) |_{\delta J_{NN}=\delta \tilde{J}_{NN}}=0$
for each $N=10,12,\dots,22$.
The least-squares fit to these data
yields an estimate 
$\rho_s/\rho_v=0.1731(50)$ in the thermodynamic limit $L\to \infty$.
The irregularity around $1/L^2=0.05(\approx 1/4.5^2)$
may be due to the artifact of the screw-boundary condition  \cite{Novotny90}. 
As in Sec. \ref{section2_1},
we consider an intermediate sector $N=14,16,\dots,20$,
for which
the least-squares fit
 leads to an estimate
$\rho_s/\rho_v=0.1820(67)$.
The discrepancy $\approx 0.009$ 
between the independent extrapolations
may provide an
indicator for a possible systematic error.
Putting both least-squares-fit and systematic errors into consideration,
we estimate the amplitude ratio as
\begin{equation}
\label{estimate2}
\rho_s/\rho_v=0.17(2)
 .
\end{equation}
The amplitude ratio
provides  a ``quantitative
measure"
\cite{Gazit14}
of
 deviation from selfduality;
we stress that
the formal duality argument does not fix 
the ratio $\rho_s / \rho_v$ quantitatively.

\subsection{\label{section2_3}
Amplitude ratios
$\rho_{s,v}/\Delta$}

In this section,
we dwell on the expressions,
$\rho_{s,v}/\Delta$,
which are the central concern in the
preceding studies.
A comparison with these studies follows.
For that purpose,
we resort to
 a relation
$m_H/\Delta =2.1(2)$ \cite{Nishiyama15}.
Based on this relation,
we convert the results,
Eqs.
(\ref{estimate1})
and 
(\ref{estimate2}),
into
\begin{equation}
\label{estimate3}
\rho_s/\Delta= 0.34(4)
 ,
\end{equation}
and
\begin{equation}
\label{estimate4}
\rho_v/\Delta = 2.0(4) .
\end{equation}

This is a good position to address an overview 
of the related studies;
see Table \ref{table1}.
According to
the Monte Carlo simulation \cite{Gazit13,Gazit14},
the
results,
 $\rho_s/\Delta=0.44(1)$, $\rho_v/\Delta=2.1(1)$ and 
$\rho_s/\rho_v=0.21(1)$, were obtained.
An elaborated non-perturbative-renormalization-group analysis
\cite{Rose17}
yields $\rho_s / \Delta=0.414$, $\rho_v/\Delta=1.98$
and $\rho_s/\rho_v=0.210$.
As a reference, we recollect pioneering renormalization-group results,
$\rho_s/\Delta=0.386$ \cite{Rose15} and 
$\rho_s/\Delta=0.2642$ \cite{Ranson13}.
The large-$N$ analysis arrives at
$\rho_v/\Delta=\frac{12}{2\pi}=1.909 \dots$
\cite{Damle97}.

As mentioned above,
the results for
$\rho_s/\Delta$
have not yet been 
 settled.
Our result
$\rho_s/\Delta=0.34(4)$,
Eq. (\ref{estimate3}), seems to support
the pioneering
renormalization-group study \cite{Rose15}.
That is, 
our result
displays a tendency toward a suppression
of  $\rho_s$,
as compared to the Monte-Carlo result.
As for the dual counterpart
$\rho_v/\Delta$,
our result 
$\rho_v/\Delta=2.0(4)$,
Eq. (\ref{estimate4}),
agrees with the
above-mentioned
preceding studies.
It would be intriguing that the 
large-$N$ result
$\rho_v/\Delta=1.909 \dots$
\cite{Damle97}
is approved by these elaborated analyses.
As to $\rho_s/\rho_v$,
our result 
$\rho_s/\rho_v=0.17(2)$,
Eq.
(\ref{estimate2}),
lies out of the above-mentioned results,
again reflecting
a
tendency toward a suppression of $\rho_s$.

\section{\label{section3}Summary and discussions}

The duality-mediated critical amplitude ratios were investigated 
for the 
$(2+1)$-dimensional $S=1$ $XY$ model 
(\ref{Hamiltonian}).
We employed
the exact diagonalization method,
which enables us to calculate the energy gap directly;
as shown in Fig. \ref{figure1},
the energy gap provides a fundamental unit for the stiffness
constants $\rho_{s,v}$.
Based on the finite-size-scaling analyses,
we obtained the estimates, 
$\rho_s/m_H=0.16(1)$, Eq. (\ref{estimate1}),
and 
$\rho_s/\rho_v=0.17(2)$, Eq. (\ref{estimate2}).
Thereby,
we converted 
these results
into 
$\rho_s/\Delta=0.34(4)$, Eq.  (\ref{estimate3}),
and
$\rho_v/\Delta=2.0(4)$, Eq.  (\ref{estimate4})
via
$m_H/\Delta =2.1(2)$ \cite{Nishiyama15}.
Our results are to be compared with
the Monte Carlo estimates
$(\rho_s/\Delta,\rho_v/\Delta,\rho_s/\rho_v)=(0.44(1),2.1(1),0.21(1))$
\cite{Gazit13,Gazit14}
and
the renormalization-group ones,
$(0.414,1.98,0.210)$ \cite{Rose17}.
It has to be mentioned that the
pioneering renormalization-group analyses
arrived at
$\rho_s/\Delta=0.386$ \cite{Rose15}
and
$\rho_s/\Delta=0.2642$ \cite{Ranson13}.
As for 
$\rho_s/\Delta$,
the results have not yet been settled.
Our result displays a tendency toward
a suppression of
 $\rho_s$, suggesting
pronounced deviation from the 
naive selfduality relation $\rho_s/\rho_v=1$.
On the one hand,
as to the dual counterpart $\rho_v/\Delta$,
the results are almost settled.
Notably enough,
the large-$N$ result 
$\rho_v/\Delta=\frac{12}{2\pi}=1.909 \dots$
is approved by the elaborated calculations.
It would be intriguing to
examine the validity of 
the $N$-dependent generic results
\cite{Rose17} systematically
through 
extending the
internal symmetries
to
$N=3,4,\dots$.

\section*{Acknowledgment}
This work was supported by a Grant-in-Aid
for Scientific Research (C)
from Japan Society for the Promotion of Science
(Grant No. 25400402).

\begin{table}
\caption{A summary of the related studies is presented.
The non-perturbative renormalization group (NPRG)
has a number of variants.
The abbreviations, DE and BMW,
denote derivative expansion and Blaizot M\'endez-Galain Wschebor,
respectively.}
\label{table1}       
\begin{tabular}{llll}
\hline\noalign{\smallskip}
Amplitude ratios & $\rho_s/\Delta$ & $\rho_v/\Delta$ & $\rho_s/\rho_v$ \\
\noalign{\smallskip}\hline\noalign{\smallskip}
This work & $0.34(4)$ & $2.0(4)$ & $0.17(2)$ \\
Monte Carlo \cite{Gazit13,Gazit14}  & $0.44(1)$ & $2.1(1)$ & $0.21(1)$ \\
NPRG-DE \cite{Rose17} & $0.414$ & $1.98$ & $0.210$ \\
NPRG-BMW \cite{Rose15} & $0.386$ & &  \\
NPRG-DE \cite{Ranson13} & $0.2642$ & & \\
\noalign{\smallskip}\hline
\end{tabular}
\end{table}

\appendix

\section*{\label{appendix}Simulation algorithm: Screw-boundary condition}

We implemented the screw-boundary condition 
\cite{Novotny90} for the two-dimensional $XY$ model 
(\ref{Hamiltonian}).
In short,
an alignment of spins $\{ {\bf S}_i \}$ ($i=1,2,\dots,N$) is arranged
so as to 
form
a
toroidal-coil structure, which is equivalent to the two-dimensional
cluster
under the screw-boundary condition.
We adopted the simulation algorithm presented in 
Ref. \cite{Nishiyama08}.
Here, for the sake of self-consistency,
we present explicit expressions 
for the perturbation operators,
$J$, $T$, and $N_k$
in the screw-boundary-condition representation.
The current operator $J$ is given by
\begin{eqnarray}
J & =& \frac{iJ_{NN}}{2} \sum_{j=1}^N (S_j^+ S_j^-(1)
                          -S_j^- S_j^+ (1))
           \nonumber \\
& & +\frac{iJ_{NNN}}{2} \sum_{j=1}^N (S_j^+ S_j^- (\sqrt{N}+1)
                          -S_j^- S_j^+ (\sqrt{N}+1))
    \nonumber \\
& & -\frac{iJ_{NNN}}{2} \sum_{j=1}^N (S_j^+ S_j^- (\sqrt{N}-1)
                          -S_j^- S_j^+ (\sqrt{N}-1))
,
\end{eqnarray}
with the $\delta$-th-neighbor spin 
operator
$S^{\pm}_{j}(\delta)=P^{\delta} S^{\pm}_j P^{-\delta}$,
and
the translation operator $P$ 
($P| \{S^z_i \} \rangle = | \{ S^z_{i+1} \} \rangle$)
\cite{Novotny90}.
The diamagnetic contribution $T$ is given by
\begin{eqnarray}
T & =& \frac{J_{NN}}{2} \sum_{j=1}^N (S_j^+ S_j^-(1)
                          +S_j^- S_j^+ (1))
           \nonumber \\
& & +\frac{J_{NNN}}{2} \sum_{j=1}^N (S_j^+ S_j^- (\sqrt{N}+1)
                          +S_j^- S_j^+ (\sqrt{N}+1))
    \nonumber \\
& & +\frac{J_{NNN}}{2} \sum_{j=1}^N (S_j^+ S_j^- (\sqrt{N}-1)
                          +S_j^- S_j^+ (\sqrt{N}-1))
.
\end{eqnarray}
Similarly, the charge-density-wave operator is defined as
$N_k= \sum_{j=1}^N e^{i k j}S^z_j$,
and the charge-density-wave susceptibility is given by
the formula
$
\chi_\rho (k) = \frac{1}{N} \langle 0 | N_k^\dagger  ({\cal H}-E_0)^{-1} N_k 
| 0 \rangle $.

%

%
%
%
%

\end{document}